\documentclass[aps,showpacs, pra, amssymb, amsmath, twocolumn]{revtex4}
\usepackage{amsmath,latexsym,amssymb}
\usepackage{graphicx}
\usepackage{amsmath}
\usepackage{latexsym}
\usepackage{amssymb}
\usepackage{bm}
\usepackage{color}
\begin{document}

\title{Quantum dynamics of a two-state system induced by a chirped zero-area pulse}

\author{Han-gyeol Lee, Yunheung Song, Hyosub Kim, Hanlae Jo, and Jaewook Ahn}
\email{jwahn@kaist.ac.kr}
\address{Department of Physics, KAIST, Daejeon 305-338, Korea}
\date{\today}

\begin{abstract}
It is well known that area pulses make Rabi oscillation and chirped pulses in the adiabatic interaction regime induce complete population inversion of a two-state system. Here we show that chirped zero-area pulses could engineer an interplay between the adiabatic evolution and Rabi-like oscillations. In a proof-of-principle experiment utilizing spectral chirping of femtosecond laser pulses with a resonant spectral hole, we demonstrate that the chirped zero-area pulses could induce, for example, complete population inversion and return of the cold rubidium atom two-state system. Experimental result agrees well with the theoretically considered overall dynamics, which could be approximately modeled to a Ramsey-like three-pulse interaction, where the $x$- and $z$- rotations are respectively driven by the hole and the main pulse. 
\end{abstract}

\maketitle

\section{Introduction}
Quantum technology, a new field of engineering based on quantum mechanics, is expected to make a defining impact on life in the 21st century~\cite{Dowling2003}. In many applications of quantum technology, such as quantum computing, quantum cryptography, quantum simulation, and quantum metrology, the ability to control the dynamics of quantum systems is an essential and foremost important necessity. Often referred to as a {\it qubit} in quantum information science~\cite{Nielson2010}, which is used as an elementary building block in more complex quantum devices and machines, a two-state quantum system is defined by a coherent superposition of two energy states. 

The dynamics of a two-state quantum system induced by coherent radiation have been extensively studied: Rabi oscillations~\cite{RabiPR, AllenBook}, Ramsey fringes~\cite{Ramsey1950}, and spin-echo dynamics~\cite{CarrPR1954} are among the best known examples. These examples are well understood in the basis of the {\it area pulse} concept~\cite{McCallPRL1967,ScullyBook}, where the pulse (temporal) area $\Theta$ is defined by 
$\Theta=\int^\infty_{-\infty}\Omega(t)dt$, 
with Rabi frequency $\Omega(t)$ that is defined ({e.g.}, in laser-atom interactions) by $\Omega(t)=\mu\mathcal{E}(t)/\hbar$ with $\mu$ the atomic transition dipole moment and $\mathcal{E}(t)$ the envelope of a laser electric-field in resonant frequency. 
Besides Rabi oscillations, there are only a few exactly solvable models, including Landau-Zener, Rosen-Zener, Allen-Eberly, Bambini-Berman, Demkov, Nikitin, and Carrol-Hioe models~\cite{Landau-Zener, Rosen-Zener, AllenBook, Bambini-Berman, Carrol-Hioe}. 

In general, coherent radiation is a complex, therefore powerful, control means; its amplitude, phase, frequency, and polarization can be used as independent control parameters.  Of particular relevance in the context of the present paper, another important control parameter in two-state system dynamics is {\it chirp}~\cite{chirp, chirpbook}, the frequency sweeping rate of coherent radiation in time. When a laser frequency $\omega(t)$ changes linearly in time with the (temporal) chirp parameter $\alpha$ as 
$\omega(t)=\omega_o+2\alpha t$,
across $\omega_o$, the Bohr transition frequency of the two-state system, a complete population inversion (CPI) between the two states occurs (when the adiabatic condition~\cite{VitanovRev} is satisfied all along the interaction), and the two-state system evolves in time through a so-called rapid adiabatic passage (RAP)~\cite{ShoreBook}.

In this paper, we consider {\it chirped zero-area pulses} to study the two-state system dynamics induced by them. To begin with, in the basis of the pulse-area theorem~\cite{ShoreBook}, ``simple'' zero-area pulses make no net transition, leaving the two-state system intact. However, to be more precise, the system does change during the dynamics, although it ends up as a complete population return (CPR) after the completion of the interaction. In that regards, certain manipulation of the initial zero-area pulse could alter the system evolution completely, leading to a significant change in the final transition probability. Even CPI could take place by detuning the zero-area pulse~\cite{VasilevPRA2006,VitanovNJP2007}, where abrupt phase change makes transitions in adiabatic basis or a crossover between adiabatic evolution and Rabi oscillations. Inspired by these counter-intuitive examples, we proceed further to consider chirping an (on-resonant) zero-area pulse in this paper. For this, we have conducted laser-atom interaction experiments with intense shaped laser pulses and spatially confined atomic vapor. The results are (1) chirped zero-area pulses could produce both CPI and CPR; (2) the excited-state probability oscillates as a function of the effective pulse-area defined by pulses with zero chirp; and (3) the given dynamics can be modeled in terms of Ramsey-type three-pulse sequence $R_x(\pi/2)R_z(\Theta_z) R_x(\pi/2)$ (to be explained below), all of which can be summarized by an interplay between adiabatic evolution and Rabi-like oscillations.
 

The rest of the paper is organized as follows: In Sec. II, we describe the model and theoretically analyze the coherent excitation induced by chirped zero-area pulses, and the detailed interpretation of the dynamics will be considered in Sec. III. Section IV is devoted to the experimental description, and we present the experimental results in Sec. V, before concluding in Sec. VI.

\section{Theoretical consideration}


When the electric field of a laser pulse is defined in time domain as
\begin{equation}
E(t) = \mathcal{E}(t) \cos{\left(\omega_o t+\varphi \right)},
\end{equation}
where $\mathcal{E}(t)$ and $\varphi$ are the envelope and phase, respectively, of the electric-field that oscillates with the carrier frequency of $\omega_o$ set to the Bohr transition frequency of the two-state atom, the pulse area $\Theta$ is given under the rotating wave approximation by 
\begin{equation}
\Theta= \frac{2\mu}{\hbar} \int^{\infty}_{-\infty}E(t)e^{-i\omega_o t} dt = \frac{\sqrt{8\pi} \mu}{\hbar} \widetilde{E}(\omega_o), \label{zeroarea}
\end{equation}
where $\widetilde{E}(\omega)$ is the amplitude spectrum of the electric field. In that regards, on-resonance pulses with zero resonant-frequency component, {\it i.e.}, $\widetilde{E}(\omega_o)=0$, are {\it zero-area pulses} ($\Theta=0$) defined in frequency domain, which is consistent with the zero-area pulse alternatively defined in time domain~\cite{VasilevPRA2006}. For instance, a Gaussian pulse with a spectral hole around the resonance frequency $\omega_o$ is a zero-area pulse, which is given by the difference of two Gaussian pulses both frequency-centered at $\omega_o$, {\it i.e.},
\begin{equation}
\frac{\widetilde{{E}}(\omega) }{E_o}=  \exp\Big[{-\frac{(\omega-\omega_o)^2}{\Delta \omega_1^2}}\Big]- \exp\Big[-\frac{(\omega-\omega_o)^2}{\Delta \omega_2^2}\Big].
\label{E_omegaz}
\end{equation}
When the spectral width of the first pulse, $\Delta\omega_1$, is significantly bigger than that of the second, $\Delta\omega_2$, ({\it i.e.}, $\Delta\omega_1 \gg\Delta\omega_2$), the second pulse (or, to say, the hole pulse) is regarded as a narrow spectral hole of the first pulse (the main pulse).

Let us consider now a {\it chirped zero-area pulse}, by chirping the above zero-area pulse, which reads
\begin{eqnarray}
\frac{\widetilde{{E}}(\omega) }{E_o} &=&  \left[\exp\Big[{-\frac{(\omega-\omega_o)^2}{\Delta \omega_1^2}}\Big]- \exp\Big[-\frac{(\omega-\omega_o)^2}{\Delta \omega_2^2}\Big]\right]  \nonumber \\
&& \times \exp\Big[ -\frac{i c_2}{2}(\omega-\omega_o)^2\Big],
\label{E_omega}
\end{eqnarray}
where the two constituent Gaussian pulses are simultaneously chirped with the frequency chirp $c_2$. The electric field in time domain is then given by
\begin{eqnarray}
E(t) &=&  \mathcal{E}_1(t) \cos{[(\omega_o +\alpha t )t + \varphi_1]} \nonumber \\
&&- \mathcal{E}_2(t) \cos{[(\omega_o + \beta t)t+ \varphi_2]} \nonumber  \\
 &\equiv&  E_1(t) - E_2(t) , 
\label{e-field}
\end{eqnarray}
where the amplitude and the phase for each pulse $i=1,2$ are, respectively, 
\begin{eqnarray}
\mathcal{E}_i(t) &=& \frac{E_o\Delta\omega_i}{\sqrt{2}} \sqrt{\frac{\tau_{o,i}}{\tau_i}} e^{-{t^2}/{\tau_i^2}}, \\
\varphi_i &=&-\frac{1}{2}\tan^{-1}\frac{2c_2}{\tau_{o,i}^2}, 
\label{echirp}
\end{eqnarray}
with $\tau_{o,i} = {2}/{\Delta\omega_i}$, the initial pulse widths, and $\tau_i=\sqrt{\tau_{o,i}^2+4c_2^2/\tau_{o,i}^2}$, the chirped pulse widths. The chirp parameters are $\alpha = {2c_2}/(\tau_{0,1}^4+4c_2^2)$ and $\beta = {2c_2}/(\tau_{0,2}^4+4c_2^2)$, respectively, for the main and hole pulses. Note that, although Eq.~\eqref{e-field} satisfies $\Theta=0$ based on the definition in Eq.~\eqref{zeroarea}, the conventional pulse area defined in time domain, $\int_{-\infty}^{\infty}\mu \mathcal{E}_{\rm total}(t)/\hbar dt$, for real envelope $\mathcal{E}_{\rm total}(t)$ of the total electric field $E(t) = E_1(t)-E_2(t)$, is generally non-zero for $c_2 \neq 0$.


As to be discussed below, this type of chirped zero-area pulses can induce both CPI and CPR of a two-state system. 
To understand the nature of the given dynamics, we will transform the two-state system in three steps: first to an interaction basis (to freeze the phase evolution of the bare atomic state), then to an adiabatic basis~\cite{VitanovAmpmod} (to trace the system's state-mixing during the interaction), and finally to another interaction basis (to clear out the remaining fast phase-evolution of the state). After obtaining the model Hamiltonian, we will discuss the resulting state-evolution in terms of Ramsey-like three rotations in the Bloch sphere representation.

\subsection{Schr\"odinger equation described in the interaction picture of the main pulse}

The dynamics of the two-state system that interacts with the electric field $E(t)$ in Eq.~\eqref{e-field} is governed by the interaction Hamiltonian $V_I$ as
\begin{eqnarray}
V_I 
&=&  \frac{\hbar}{2} \left[ \begin{array}{cc}
-\Delta(t) & \Omega_1(t) \\
\Omega_1(t) & \Delta(t)
\end{array} \right] 
-\frac{\hbar}{2}
\left[ \begin{array}{cc}
0 & {\Omega}_2(t) \\
{\Omega}_2^*(t)  & 0
\end{array} \right] \nonumber \\
&\equiv& V_{I,1}+V_{I,2},
\label{Hamiltonian3}
\end{eqnarray}
where $\Delta(t)=\omega_o-\omega(t)=-2\alpha t$ is the (instantaneous) detuning~\cite{Baumert}. The Rabi frequencies of the main and hole pulses are, respectively, given by
\begin{eqnarray}
\Omega_1(t) &=& \frac{\mu }{\hbar} \mathcal{E}_1(t), \\
{\Omega}_2(t)&=&\frac{\mu }{\hbar} \mathcal{E}_2(t)  e^{{i}[(\beta-\alpha)t^2+\varphi]},
\label{omegat}
\end{eqnarray}
where we define $\varphi\equiv \varphi_1-\varphi_2$, and the time-independent phase $\varphi_1$ is included in the base vector for the sake of simplicity.

\subsection{Dynamics described in the adiabatic basis of the main pulse}

In Eq.~\eqref{Hamiltonian3}, $V_{I,1}$, the first part of the interaction Hamiltonian, varies slowly compared to the second $V_{I,2}$, so the system dynamics can be more easily understood in the adiabatic basis~\cite{VitanovAmpmod} of the main pulse. The eigenvalues of $V_{I,1}$ are given by
\begin{equation}
\frac{\hbar}{2}\lambda_\pm(t) = \pm \frac{\hbar}{2}\sqrt{\Omega_1^2(t)+\Delta^2(t)}
\label{eq3}
\end{equation}
and the corresponding eigenstates are
\begin{eqnarray}
|\psi_-(t)\rangle &=& \cos\vartheta(t)|0\rangle_I -\sin\vartheta(t)|1\rangle_I, \nonumber \\
|\psi_+(t)\rangle &=& \sin\vartheta(t)|0\rangle_I +\cos\vartheta(t)|1\rangle_I
\label{eq4},
\end{eqnarray}
where $|0\rangle_I$ and $|1\rangle_I$ form the eigenbasis in the interaction picture, 
and the mixing angle $\vartheta(t)$ is defined by
\begin{equation}
\vartheta(t) = \frac{1}{2} \tan^{-1} \frac{\Omega_1(t)}{\Delta(t)} \quad \rm{for} \quad 0 \le\vartheta(t)\le\frac{\pi}{2}.
\label{mixingangle}
\end{equation}
So, the two-state system can be described in adiabatic basis ($|\psi_-(t)\rangle, |\psi_+(t)\rangle$) with the transformation $|\psi(t)\rangle_A=R(\vartheta)|\psi(t)\rangle_I$, where the rotation $R(\vartheta)$ is defined by
\begin{equation}
R(\vartheta)=\left[ 
\begin{array}{cc}
\cos\vartheta(t) & -\sin\vartheta(t) \\
\sin\vartheta(t) & \cos\vartheta(t)
\end{array} 
\right].
\end{equation}
Note that each adiabatic base vector, $|\psi_-(t)\rangle$ or $|\psi_+(t)\rangle$, changes from one atomic state to the other, as time evolves from $t=-\infty$ to $\infty$. In other words, when $c_2>0$,
\begin{eqnarray}
\lim_{t \rightarrow -\infty} |\psi_-(t)\rangle = |0\rangle_I, &\quad&
\lim_{t \rightarrow \infty} |\psi_-(t)\rangle = -|1\rangle_I, \nonumber \\
\lim_{t \rightarrow -\infty} |\psi_+(t)\rangle = |1\rangle_I, &\quad& 
\lim_{t \rightarrow \infty} |\psi_+(t)\rangle = |0\rangle_I,
\end{eqnarray}
and when $c_2<0$, the relation is reversed, since the time dependence of $\Delta$ is opposite.

In the given adiabatic basis, the  Schr\"odinger equation is given by
\begin{eqnarray}
i\hbar \frac{d }{dt} |\psi(t)\rangle_A = \left( R V_I R^{-1}  +  i\hbar R \dot{R}^{-1}  \right) |\psi(t)\rangle_A,
\label{S_equation}
\end{eqnarray}
where the second term in the parenthesis can be ignored, when the evolution by $\Omega_1(t)$ is adiabatic, {\it i.e.},
\begin{equation}
i\hbar R\dot{R}^{-1}
= i\hbar \left[
\begin{array}{cc}
0 & -\dot{\vartheta} \\
\dot{\vartheta} & 0
\end{array} 
\right] \approx 0,
\end{equation}
so the resulting interaction Hamiltonian is given  by
\begin{equation}
\label{VA}
V_{A} 
= \frac{\hbar}{2} \left[
 \begin{array}{cc}
\lambda_- & -\Omega_2  \\
-\Omega^*_2  & \lambda_+
\end{array} 
\right] 
+ \hbar {\Re}(\Omega_2) \sin\vartheta \left[
 \begin{array}{cc}
\cos\vartheta & \sin\vartheta   \\
\sin\vartheta   & -\cos\vartheta 
\end{array} 
\right],  
\end{equation}
where $\Re({\Omega}_2)$ 
is the real part of $\Omega_2(t)$. 

\subsection{Ramsey-type three pulsed interactions}

Let us take a closer look at Hamiltonian $V_A$.  At the extreme times $|t|\rightarrow \infty$, it becomes
\begin{eqnarray}
V_A & = &  \frac{\hbar}{2} \left[ \begin{array}{cc}
-|\Delta(t)| & -{\Omega}_2(t) \\
-\Omega^*_2(t) & |\Delta(t)|
\end{array} 
\right]  \quad {\rm for} \quad t \rightarrow -\infty \nonumber \\
&=& \frac{\hbar}{2} \left[ \begin{array}{cc}
-|\Delta(t)| & \Omega_2^*(t) \\
\Omega_2(t) & |\Delta(t)|
\end{array} 
\right]  \quad\;\; {\rm for} \quad t \rightarrow \infty.
\end{eqnarray}
Note that the phase evolutions of the diagonal terms are opposite with each other. So, it is convenient to remove the phase factor, $\exp(i\int_0^t|\Delta(t')|dt') = \exp(i|\Delta(t)|t/2)$, through a transformation to another interaction basis, {\it i.e.},
\begin{equation}
|\psi(t)\rangle_F \equiv T_\Delta|\psi(t)\rangle_A,
\label{transform}
\end{equation}
where $T_\Delta= \exp\Big({i} \int_{0}^{t} H_\Delta(t') dt'/{\hbar} \Big)$ with
\begin{equation}
H_\Delta = \frac{\hbar}{2}\left[ \begin{array}{cc}
-|\Delta(t)| & 0  \\
0 & |\Delta(t)|
\end{array} \right].
\end{equation}
The resulting final Hamiltonian becomes
\begin{equation}
V_{F} = T_\Delta(V_A-H_\Delta)T_\Delta^{\dagger}
=  \frac{\hbar}{2} \left[ \begin{array}{cc}
-\Delta_F(t) & \Omega_F(t) \\
\Omega_F^*(t) & \Delta_F(t)
\end{array} \right], 
\label{HF}
\end{equation}
where the effective ``detuning'' $\Delta_F$ and ``coupling'' $\Omega_F$ are given by
\begin{eqnarray}
\Delta_F&=&\sqrt{\Omega_1^2(t)+\Delta^2(t)}+\Re(\Omega_2)\sin[2\vartheta(t)]-|\Delta(t)|,  \nonumber \\
\Omega_F 
&=& \{-\cos[2\vartheta(t)]\Re(\Omega_2)-i\Im(\Omega_2)\}e^{-i {|\Delta(t)|t}/{2}}
\label{omegaf},
\end{eqnarray} 
where $\Im({\Omega}_2)$ is the real part of $\Omega_2(t)$. The above Hamiltonian leads to the detuned Rabi oscillation~\cite{ShoreBook} after a second rotating-wave approximation. The remaining time-dependence in $\Omega_F(t)$, the slowly-varying phase factor $e^{i\beta t^2}$, is negligible for the considered chirp values.

\begin{figure}
    \centerline{\includegraphics[width=0.45\textwidth]{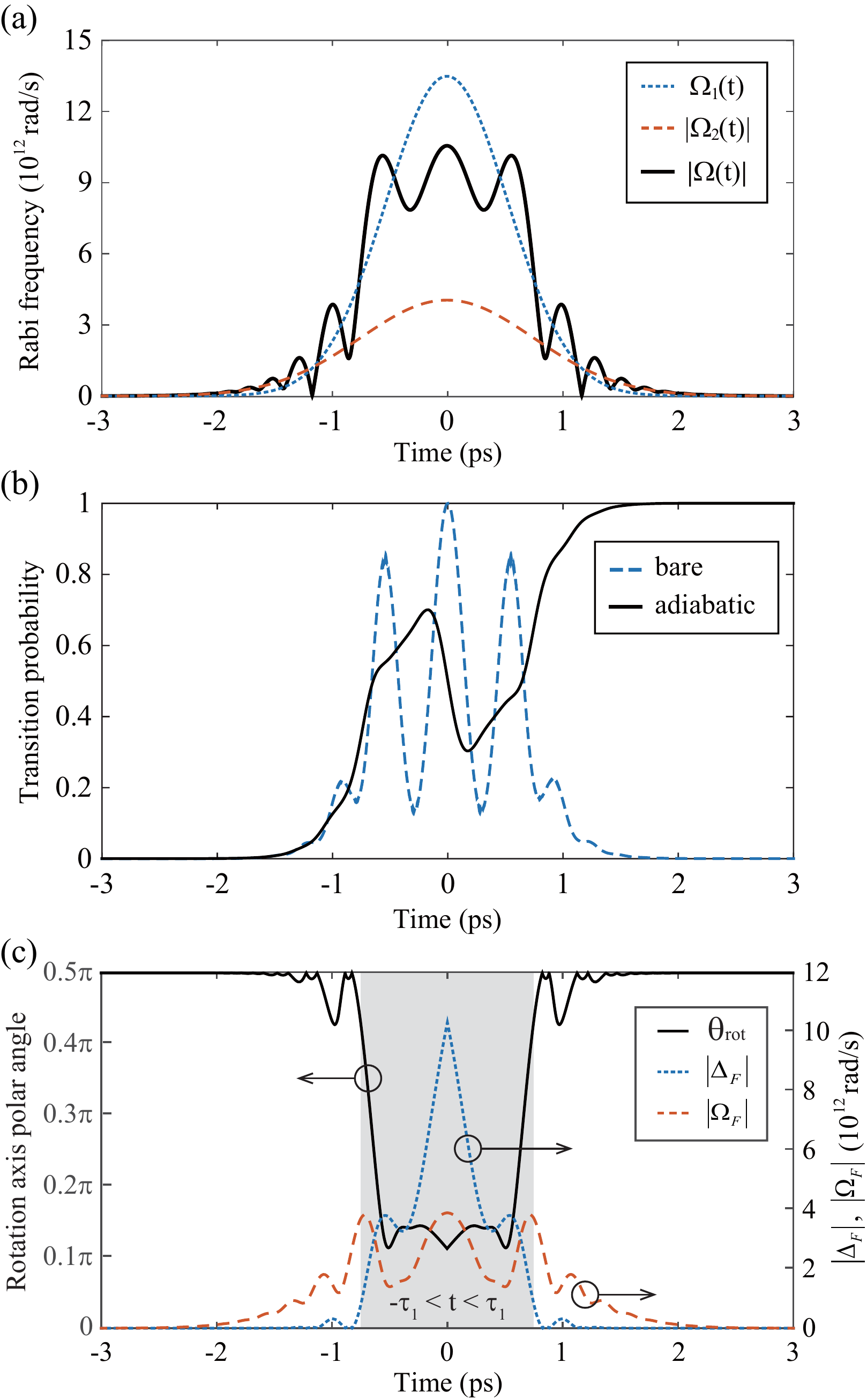}}
    \caption{(Color online) (a) The Rabi frequencies $\Omega_1(t)$ (dotted line) for the main pulse,  $|\Omega_2(t)|$ (dashed line) for the hole, and $|\Omega(t)|$ (solid line) for the total electric field. (b) The time evolution of the transition probabilities in bare atom and adiabatic bases.  (c) The transience of the polar angle $\theta_{\rm{rot}}(t)$ (solid line) for  the rotational axis of the Bloch vector in the adiabatic basis compared with $|\Delta_F(t)|$ (dotted line) and $|\Omega_F(t)|$ (dashed line). The calculation parameters chosen at a CPR are given in the text. }\label{FigPulse}
\end{figure}

Figure~\ref{FigPulse}(a) plots the Rabi frequencies of the main pulse, $\Omega_1(t)$,  the hole, $|\Omega_2(t)|$, and the total electric field, $|\Omega(t)|=|\Omega_1(t)-\Omega_2(t)|$, of a chirped zero-area pulse that results in a CPR.  The numerical values for the given pulse are $c_2=5.1\times 10^4$~fs$^2$,  $\Delta \omega_1= 1.5\times 10^{13}$~rad/s, and $\Delta\omega_2=1.9 \times 10^{12}$~rad/s. The peak electric field of the main pulse is $E_o\Delta\omega_1/\sqrt{2}=2.3\times10^8$~V/m, and the dipole moment of atomic rubidium ($^{85}$Rb) for linearly polarized light is given by $\mu= 1.46 \times 10^{-29}$~Cm ~\cite{KimPRA2015, Rbdata}. To examine adiabaticity, we use the adiabaticity function $f(t) = {|\dot{\Omega}(t)\Delta(t)-\Omega(t)\dot{\Delta}(t)|}/{2[\sqrt{\Omega^2+\Delta^2}]^3}$~\cite{VitanovRev}. Adiabatic evolution condition is given by $f(t)\ll1$. For linearly chirped pulses, the function $f(t)$ can be explicitly written as,
\begin{eqnarray}
f_1(t)&=& \frac{\alpha|\Omega_1(t)|(2t^2/\tau_1^2+1)}{[\sqrt{|\Omega_1|^2 + 4\alpha^2t^2}]^3}, \\
f_2(t)&=& \frac{\beta|\Omega_2(t)|(2t^2/\tau_2^2+1)}{[\sqrt{|\Omega_2|^2 + 4\beta^2t^2}]^3},
\end{eqnarray}
where $f_1(t)$ and $f_2(t)$ are the adiabatic functions for the main and the hole pulses, respectively. Note that the pulse width $\tau_i$ and the chirp parameters $\alpha$ and $\beta$ are functions of the bandwidth $\Delta\omega_i$ as well as of the linear chirp $c_2$. 
With the given parameters in Fig.~\ref{FigPulse}, we obtain the adiabatic condition, $f_1(t)<0.2$, for the main pulse with $E_o\Delta\omega_1/\sqrt{2}>1.2\times 10^8$~V/m and $c_2>2\times 10^4$~fs$^2$, which ensures the adiabatic evolution by the main pulse. However, we obtain $f_2(t)>1.2$, so the hole pulse induces simple Rabi oscillations, even up to the region $c_2\simeq1\times10^5$~fs$^2$. 

In Fig.~\ref{FigPulse}(b), the time-evolution of the transition probability in adiabatic basis is compared with the transition in the bare basis, which clearly shows  the transition in the adiabatic basis, or a CPR.  Also, we plot in Fig.~\ref{FigPulse}(c) the polar angle, $\theta_{\mathrm{rot}}(t)$, of the rotational axis on the Bloch sphere in the adiabatic basis, defined by 
\begin{equation}
\theta_{\mathrm{rot}}(t) = \tan^{-1}  \frac{|\Omega_F(t)|}{|\Delta_F(t)|}.
\label{rotangle}
\end{equation}
The rotation on a Bloch sphere is approximately a phase evolution (about the $z$ axis) when $\theta_{\mathrm{rot}} = 0$, and Rabi oscillations when $\theta_{\mathrm{rot}} = 0.5\pi$. Therefore, the plateaus in Fig.~\ref{FigPulse}(b) suggest that there exist three distinct coupling regimes:  In the two tail regions ($t<-\tau_1$ and $t>\tau_1$), the ``coupling'' is dominant, so a Rabi oscillation is expected; and, in the central region ($-\tau_1<t<\tau_1$), the ``detuning'' is large, effectively causing approximately a phase evolution of the system.

\subsubsection{$|\Delta(t)| \gg \Omega_1(t)$ (the tail regions)}

In the tail regions ($t < -\tau_1$ and $t > \tau_1$), the detuning $\Delta(t)$ greatly exceeds the main-pulse interaction $\Omega_1(t)$, {\it i.e.}, $|\Delta(t)| \gg \Omega_1(t)$. So, we get $\Delta_F \rightarrow 0$, and $\Omega_F \rightarrow -|\Omega_2|e^{i\varphi}$ (for $t<\tau_1$) or $|\Omega_2|e^{-i\varphi}$ (for $t>\tau_1$). Given that, the Hamiltonian in Eq.~\eqref{HF} is approximated, when the slowly varying phase $e^{i\beta t^2}$ in $\Omega_2$ is neglected, as
\begin{eqnarray}
V_{F} &\approx&  \frac{\hbar}{2} \left[ \begin{array}{cc}
0 & {|{\Omega}_2(t)|} e^{i(\pi+\varphi)} \\
{|{\Omega}_2(t)|} e^{-i(\varphi+\pi)} & 0
\end{array} \right] \; {\rm for} \; t<-\tau_1 \nonumber \\
&\approx&  \frac{\hbar}{2} \left[ \begin{array}{cc}
0 & {|{\Omega}_2(t)|} e^{-i\varphi} \\
{|{\Omega}_2(t)|} e^{i\varphi} & 0
\end{array} \right] \; {\rm for} \; t>\tau_1,
\label{Hamiltonian2nd}
\end{eqnarray}
which means that the dynamics in the tail regions are rotations on the Bloch-sphere surface, represented by $R_{\pi+\varphi}(\Theta_2^-)$ and $R_{-\varphi}(\Theta_2^+)$,  respectively, for $t<\tau_1$ and $t>\tau_1$. The rotation axes are on the $xy$ plane, of which the directions are defined by azimuthal angles $\phi=\pi+\varphi$ and $\phi=-\varphi$, respectively. The rotation angles are given by 
\begin{equation}
\Theta_2^-=\int_{-\infty}^{-\tau_1}|{\Omega}_2(t)|dt  \quad \mbox{and} \quad
\Theta_2^+=\int^{\infty}_{\tau_1}|{\Omega}_2(t)|dt,
\label{theta2}
\end{equation}
as a function of the hole pulse $E_2$(t), and $\Theta_2^+$=$\Theta_2^-$ due to the symmetry.

\subsubsection{$|\Delta(t)| \ll \Omega_1(t)$ (the central region)}

In the central time region ($-\tau_1<t<\tau_1$), the main pulse $E_1(t)$ plays an important role. To understand the dynamics in this region, we consider an extreme approximation under the conditions  $\Omega_1(t) \gg \Delta(t)$ and $\Omega_1(t) \gg |\Omega_2(t)|$. (Note that the case depicted in Fig. 1 is different from this extreme case.) Then, we get $\Delta_F\approx \Omega_1(t)-|\Delta(t)|$, which leads to 
\begin{equation}
V_F  \approx \frac{\hbar}{2} \left[
\begin{array}{cc}
-\Omega_1(t)+|\Delta(t)| & 0 \\
0 & \Omega_1(t)-|\Delta(t)| 
\end{array} 
\right].
\end{equation}
The resulting two-state dynamics is the rotation about the $z$-axis, or $R_z(\Theta_1)$, by an angle 
\begin{equation}
\Theta_1=\int^{\tau_1}_{-\tau_1} ( \Omega_1(t)-|\Delta(t)|)dt
\label{theta1}
\end{equation}
defined by the main pulse $E_1(t)$.
\newline

As a result, the overall dynamics of the two-state system by the chirped zero-area pulse (in the transformed adiabatic basis) can be summarized as following:
\begin{eqnarray}
R_{-\varphi}(\Theta_2^+)R_z(\Theta_1) R_{\pi+\varphi}(\Theta_2^-).
\label{RRR},
\end{eqnarray}
which is not in the bare basis. This manifestation of the three-step interaction is a reminiscence of the Ramsey rotation often described by $R_x(\pi/2)R_z(\Theta_z) R_x(\pi/2)$, from which the difference is to be discussed.

Note that the necessary condition required (in our consideration) for Ramsey-type three pulse sequence (First Rabi + Second adiabatic evolution + Third Rabi) modeling is in fact the contrast in the spectral width between the main and hole pulses. When the spectral width of the main pulse is large enough, chirping makes the main pulse satisfy the adiabatic condition, while the same amount of chirp makes the spectrally-narrow hole pulse remain as a Rabi-inducing pulse. Therefore, pulses of smooth envelope shapes other than Gaussian, when they have the spectral hole and the chirp, could also be approximated to a Ramsey-type three pulse sequence.

\section{Calculation results}

\begin{figure}
    \centerline{\includegraphics[width=0.45\textwidth]{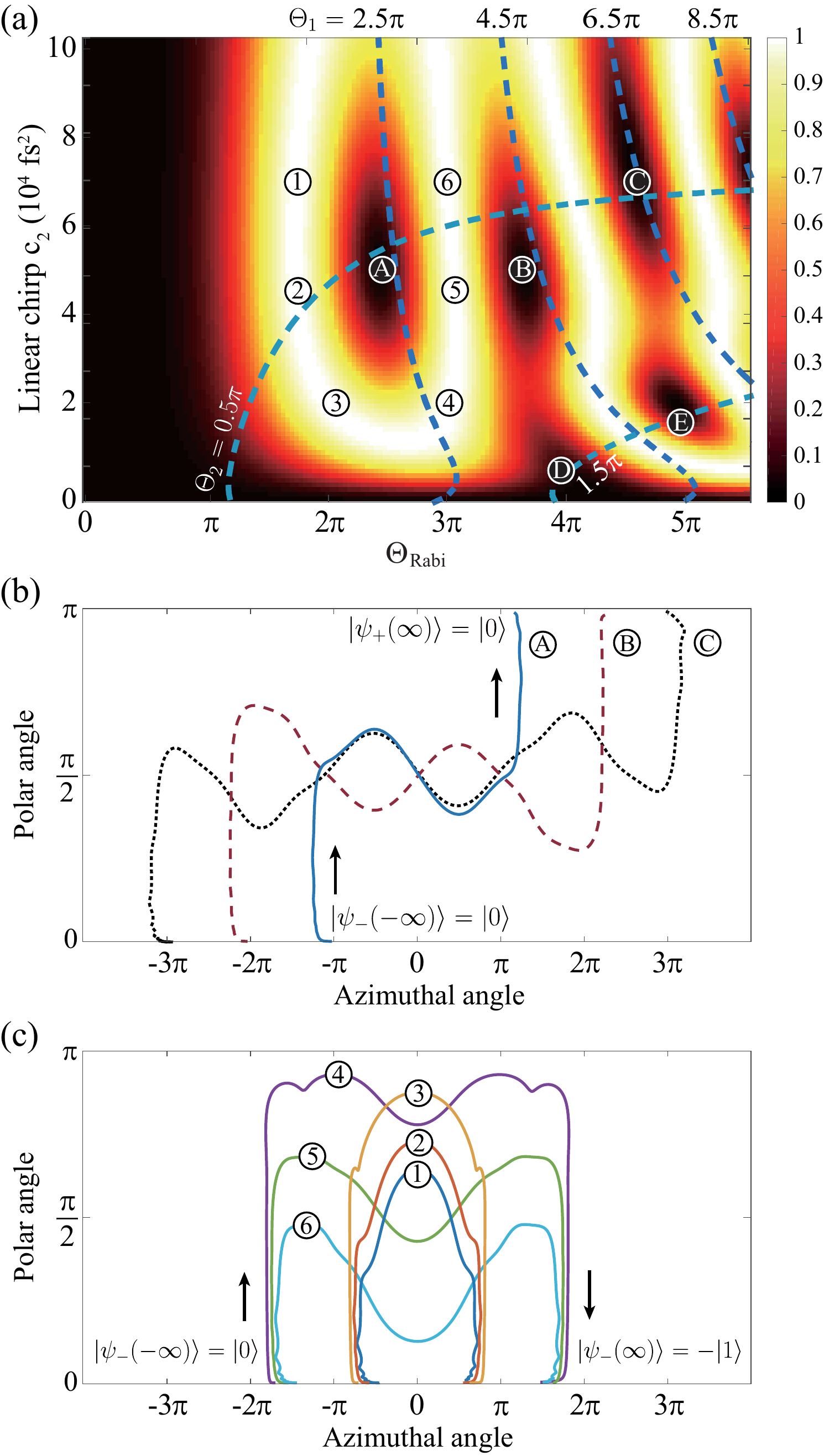}}
    \caption{(Color online) (a) Numerical calculation of the transition probability in the atomic basis is plotted as a function of $\Theta_{\rm Rabi}$, the Rabi phase of pulses with zero chirp, and $c_2$, the frequency chirp. Localized CPR regions appear as spots around \textcircled{A}, \textcircled{B}, $\cdots$, \textcircled{E}; and CPI regions as strips, {e.g.}, along \textcircled{1}-\textcircled{6}.  The dotted lines indicate contours for the estimated rotation angles $\Theta_1$ and $\Theta_2$ defined by Eqs.~\eqref{theta2} and \eqref{theta1}. (b) and (c) Trajectories of CPR and CPI represented in the adiabatic basis for chosen points in (a), respectively. 
  } \label{FigP3}
\end{figure}

Figure~\ref{FigP3} presents the numerical simulation for the two-state system dynamics induced by the chirped zero-area pulses. 
For the numerical calculation, the Schr\"{o}dinger equation in Eq.~\eqref{S_equation} is used with $\Delta \omega_1= 1.5\times 10^{13}$~rad/s and $\Delta\omega_2=1.9 \times 10^{12}$~rad/s.
In Fig.~\ref{FigP3}(a), the probability of the transition to the excited state in bare atomic basis is shown as a function of $\Theta_{\rm Rabi}$ and $c_2$, where $\Theta_{\rm Rabi}$ denotes the effective Rabi oscillation phase for the main pulse with zero chirp, defined by
\begin{eqnarray}
\Theta_{\mathrm{Rabi}} = \frac{\mu}{\hbar}\int_{-\infty}^{\infty}\mathcal{E}_{1}(t;c_2=0)dt.
\label{trabi3}
\end{eqnarray}
Note that $\Theta_{\mathrm{Rabi}}$ was varied by changing $E_o$, the electric field amplitude, in experiment and calculation.

We can identify several CPR and CPI regions in Fig.~\ref{FigP3}:  The CPR regions appear as localized dark spots, besides the wide dark region in low pulse energies; and the CPI regions are bright strips. To understand the dynamics, we select a few characteristic points in Fig.~\ref{FigP3}(a) and trace their quantum trajectories in adiabatic basis. In Figs.~\ref{FigP3}(b,c), the trajectories are plotted on the $(\theta, \phi$) surface of the Bloch sphere (defined in adiabatic basis), where $\theta$ and $\phi$ denote the polar and azimuthal angles, respectively, of the Bloch vector. At the CPR points, marked by \raisebox{.5pt}{\textcircled{\raisebox{-.9pt} {A}}}, 
\raisebox{.5pt}{\textcircled{\raisebox{-.7pt} {B}}}, and
\raisebox{.5pt}{\textcircled{\raisebox{-.9pt} {C}}}, each trajectory starts from $\theta=0$ and ends at $\theta=\pi$, as shown in Fig.~\ref{FigP3}(b). Likewise, at the CPI points, marked by \raisebox{.5pt}{\textcircled{\raisebox{-.9pt} {1}}}, 
\raisebox{.5pt}{\textcircled{\raisebox{-.7pt} {2}}}, $\cdots$, and
\raisebox{.5pt}{\textcircled{\raisebox{-.9pt} {6}}}, each trajectory starts from $\theta=0$ and ends at $\theta=0$, as shown in Fig.~\ref{FigP3}(c). Note that CPR in bare atomic basis, $|0\rangle  \to |0\rangle$, is CPI in adiabatic basis, $|\psi_-(-\infty)\rangle \to |\psi_+(\infty)\rangle$, and CPI in bare atomic basis, $|0\rangle  \to -|1\rangle$, is CPR in  adiabatic basis, $|\psi_-(-\infty)\rangle \to |\psi_-(\infty)\rangle$.

In particular, each CPR trajectory (CPI in adiabatic basis) in Fig.~\ref{FigP3}(b) consists of three distinctive rotations: a rotation of $\Delta\theta=\pi/2$ about a rotation axis in the $xy$-plane, a wobbling rotation about the $z$-axis, and again a $\Delta\theta=\pi/2$ rotation about another axis in the $xy$-plane. Note that the azimuthal rotation angles for the second wobbling rotations for the \raisebox{.5pt}{\textcircled{\raisebox{-.9pt} {A}}}, 
\raisebox{.5pt}{\textcircled{\raisebox{-.7pt} {B}}}, 
\raisebox{.5pt}{\textcircled{\raisebox{-.9pt} {C}}} trajectories are about $\Delta\phi=2\pi$, $4\pi$, and $6\pi$, respectively.  The CPR trajectories are therefore represented in accordance with Eq.~\eqref{RRR} by 
\begin{equation}
R_{-\varphi}(\pi/2)R_z(2n\pi+\gamma) R_{\pi+\varphi}(\pi/2),
\label{RRRgamma}
\end{equation}
where $n$ is a positive integer. It is noted that $\gamma=-\varphi-(\pi+\varphi)=\pi-2\varphi$ compensates the angle difference between the first and third rotation axes, explaining the deviation of the azimuthal rotation angle from $2n\pi$. Here, from Eqs.~\eqref{echirp} and \eqref{omegat}, $\varphi$ is given by $\varphi = \varphi_1-\varphi_2 = [\tan^{-1}(2c_2/\tau_{0,1}^2)-\tan^{-1}(2c_2/\tau_{0,2}^2)]/2$. Since $\Delta\omega_1>\Delta\omega_2$, $\gamma$ varies from $\pi$ to $\sim$$\pi/2$, while $c_2$ varies from 0~fs$^2$ to 100000~fs$^2$, and $\gamma\simeq 0.6\pi$ at \raisebox{.5pt}{\textcircled{\raisebox{-.9pt} {A}}}, \raisebox{.5pt}{\textcircled{\raisebox{-.7pt} {B}}}, and \raisebox{.5pt}{\textcircled{\raisebox{-.9pt} {C}}}.
Similarly, the trajectories for \raisebox{.5pt}{\textcircled{\raisebox{-.9pt} {D}}}  \raisebox{.5pt} and {\textcircled{\raisebox{-.9pt} {E}}} (not shown) can be approximately represented by $R_{-\varphi}(3\pi/2)R_z(2\pi+\gamma) R_{\pi+\varphi}(3\pi/2)$ and $R_{-\varphi}(3\pi/2)R_z(4\pi+\gamma) R_{\pi+\varphi}(3\pi/2)$, respectively. 

The actual trajectories in Fig.~\ref{FigP3}(b) are slightly deviated from the approximated model in Eq.~\eqref{RRR}. It can be explained that, in the central region, the rotation axis is not perfectly aligned with the $z$-axis, causing the wobbling of the $z$-rotation. As a result, CPRs do not occur at exact $\Theta_2 = n\pi+\pi/2$. Also, in the tail regions, the mixing angle $\vartheta(t)$ does not converge to 0 (for $t<-\tau_1$) or $\pi/2$ (for $t>\tau_1$) during the evolution, and gives $\cos[2\vartheta(t)]=1-\delta$ (for $t<-\tau_1$) and $\cos[2\vartheta(t)]=-1+\delta$ (for $t>\tau_1$) in Eq.~\eqref{omegaf}, where $\delta>0$ is a small deviation. Since the azimuthal angle of a rotation axis is given by $\arg[\Omega_F(t)]$, the actual azimuthal angle difference between the two rotation axes becomes smaller than the model. Thus, as shown in Fig.~\ref{FigP3}(a), the CPRs (\raisebox{.5pt}{\textcircled{\raisebox{-.9pt} {A}}}, \raisebox{.5pt}{\textcircled{\raisebox{-.7pt} {B}}}, and \raisebox{.5pt}{\textcircled{\raisebox{-.9pt} {C}}}) occur near $\Theta_1 = 2n\pi + 0.5\pi$, rather than $\Theta_1 = 2n\pi + \gamma~(\simeq2n\pi+0.6\pi)$. 

The intermediate region between the tail regions and the central region, appearing in Figs.~\ref{FigP3}(b,c) around the end the vertical evolution and beginning of the wobbling $z$-rotation, is short in time compared to other regions (the ratio is roughly 0.2 compared to the central region). Also, the change of the rotation axis is much faster (roughly 5 times faster) than the rotation. Therefore, as shown in Figs.~\ref{FigP3}(b,c), there are no significant influence to the dynamics in the intermediate region. 

\section{Experimental Setup}
\begin{figure}
    \centerline{\includegraphics[width=0.45\textwidth]{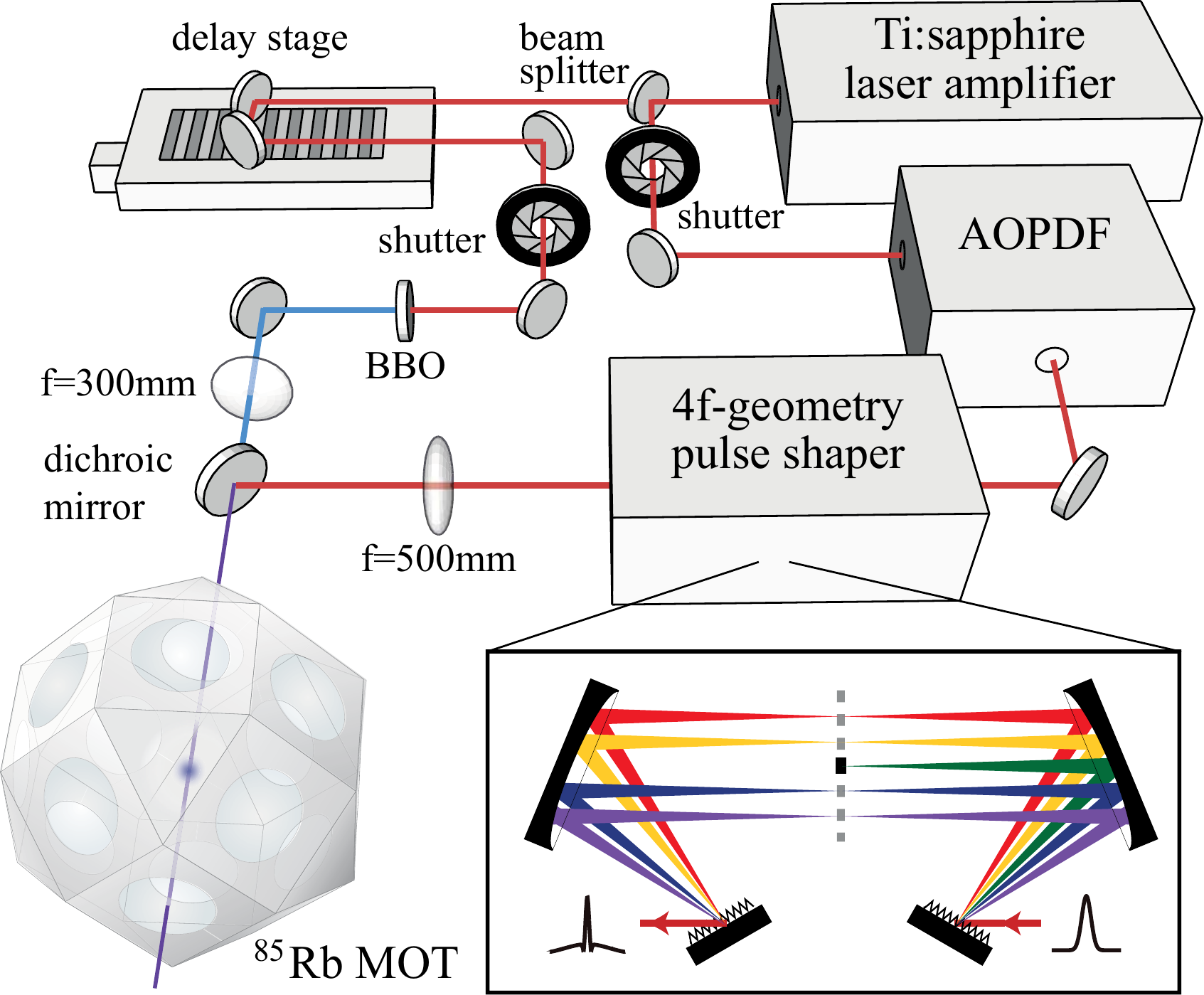}}
    \caption{(Color online) Schematics of the experimental setup. Femto-second laser pulses were shaped by an acousto-optic programmable dispersive filter (AOPDF) and a $4f$-geometry zero-dispersion stretcher; and as-produced chirped zero-area pulses interacted with cold rubidium atoms ($^{85}$Rb) in a magneto-optical trap. After 3~ns, frequency-doubled laser pulses ionized the excited atoms. 
  } \label{FigSetup}
\end{figure}

The experimental investigation of the chirped-zero area-pulse interaction of the two-state system is performed with shaped intense laser pulses and cold atomic vapor. The experimental setup is shown in Fig.~\ref{FigSetup}. Femto-second laser pulses were first produced from a home-made Ti:sapphire laser oscillator with 80~MHz repetition rate, and amplified about $5\times10^5$ times by a home-made Ti:sapphire multi-pass amplifier operating at 1~kHz repetition rate. The wavelength of the pulses was centered at $\lambda_o=794.7$~nm, resonant to the atomic rubidium ($^{85}$Rb) 5S$_{1/2}$ $\to$ 5P$_{1/2}$ transition. The two states of the quantum system were $|0\rangle=|$5S$_{1/2}\rangle$ and $|1\rangle=|$5P$_{1/2}\rangle$. The laser bandwidth was $\Delta \lambda_{\rm FWHM}=7$~nm ($\Delta \omega_1 = 1.5\times 10^{13}$~rad/s), which was narrow enough to restrict the quantum system to a two-state system, and the corresponding pulse duration was $\tau_{0,1} ~\simeq 140$~fs. Since the hyperfine splittings of $|$5S$_{1/2}\rangle$ and $|$5P$_{1/2}\rangle$ states (up to 3GHz) are much smaller than the laser bandwidth of THz order, the hyperfine states can be reduced to a two-state system through Morris-shore transform~\cite{KimPRA2015}, giving $\mu= 1.46 \times 10^{-29}$~Cm~\cite{Rbdata}.

Each femto-second laser pulse was divided into two: the first (control) pulse after being shaped to a chirped zero-area pulse induced the atomic transition, and the second (probe) pulse was frequency-doubled to ionize atoms in the excited state. The chirped zero-area pulse was prepared in two stages: the laser pulse was first frequency-chirped, up to $c_2=6\times 10^4$~fs$^2$, by an acousto-optic programmable dispersive filter (Dazzler from Fastlite)~\cite{AOPDF}, and the spectrum near the resonance was removed by a spectral block in the Fourier plane defined by a $4f$-geometry Martinez zero-dispersion stretcher (homemade)~\cite{MartinezOL1984, WeinerOC2011}, which consisted of a pair of R=500~mm cylindrical mirrors and a pair of gratings with 1800~grooves/mm. For the spectral block, we used various metal wires of which the spectral width ranges from 200 to 1000~GHz in FWHM (or $\Delta \omega_2=7.7\times 10^{11} \sim 3.9\times 10^{12}$~rad/s in Gaussian width). After the two-stage of pulse shaping, the laser pulse energy was up to 20~$\mu$J.

We used a conventional magneto-optical trap (MOT) to spatially isolate the rubidium atoms~\cite{LimSR2014}. By adjusting the diameters of the cooling and repumping laser beams, an atomic cloud of about 200~$\mu$m diameter and $6\times 10^9$~cm$^3$ atom density  was prepared. The atoms were tightly confined in particular to achieve a uniform laser-atom interaction~\cite{LeeOL2015}. When the laser pulses were focused on to the atomic vapor, the laser beam diameter of 600~$\mu$m was about 3 times bigger than the diameter of the atom cloud. With this diameter ratio, we achieved 95\% of high fidelity for a $\pi$-area transform-limited pulse excitation. The laser pulse energy (up to 20~$\mu J$) with the given beam diameter was equivalent to $\Theta_{\rm max}=3.5\pi$. After being interacted with the control pulse (the chirped zero-area pulse), atoms in the excited state were ionized by the probe pulse, and resulting ions were measured by a multichannel plate detector. The overall experimental cycle controlled by mechanical shutters in laser beam lines was 2~Hz to grant the restoration of the MOT.

\section{Experimental Results and discussion}
\label{Results}

\begin{figure}
    \centerline{\includegraphics[width=0.45\textwidth]{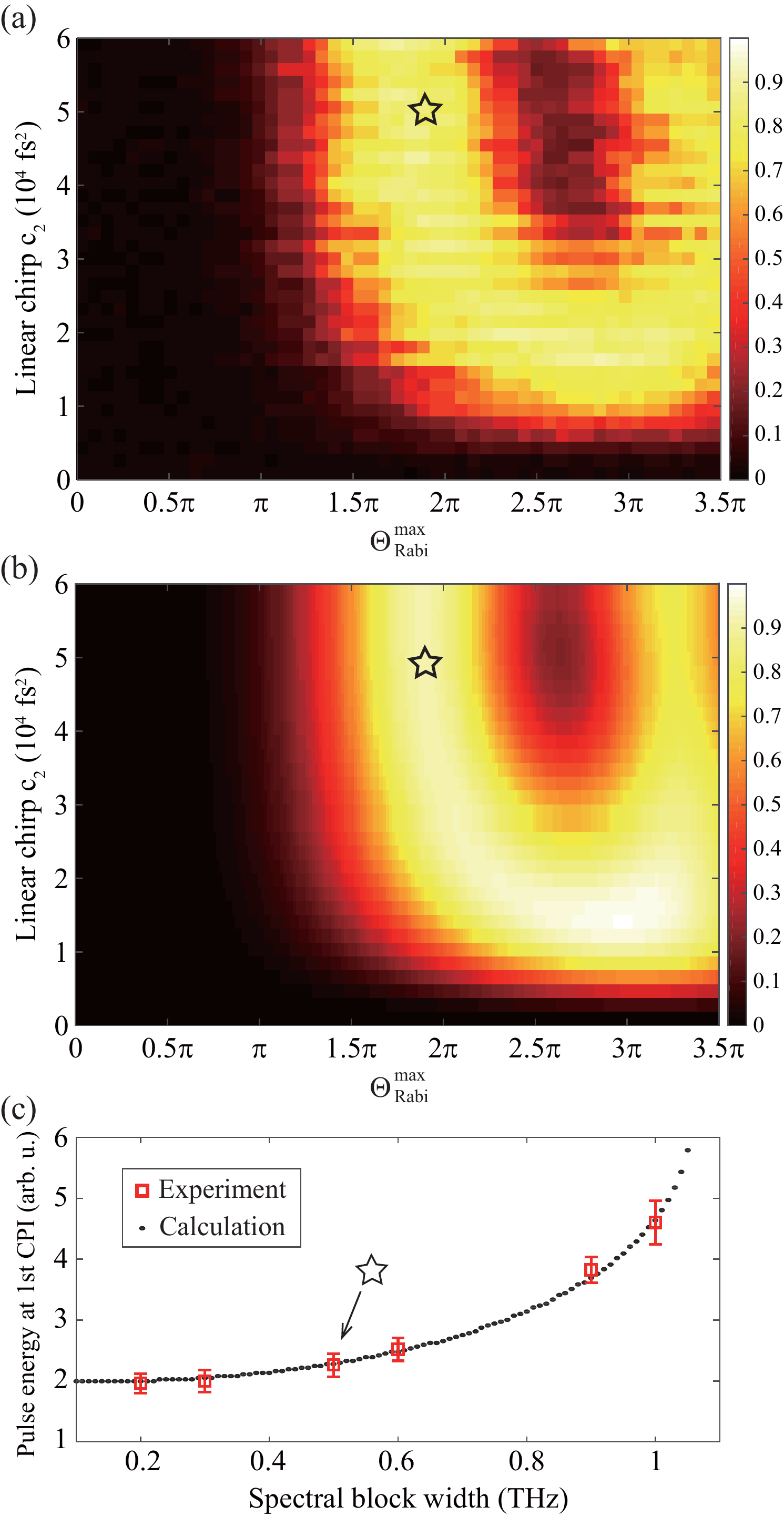}}
    \caption{(Color online) (a) Experimental result and (b) numerical simulation of the chirped zero-area pulse excitation of a cold rubidium atom ensemble. The transition probabilities were plotted as a function of $c_2$ (linear chirp) and $\Theta_{\mathrm{Rabi}}^{\rm max}$. Interaction inhomogeneity due to the Gaussian spatial profile of the atom ensemble was taken into account for the calculation of (b). (c) Experimental result of spectral hole-width scan. Measured pulse energies for the first CPI ({e.g.}, the stars marked in (a) and (b)) are plotted for various spectral hole widths (red squares), in comparison with the calculation result (black dots).
  } \label{FigQ}
\end{figure}

Figure~\ref{FigQ}(a) shows the experimental result. By varying the power and the chirp parameter $c_2$ of the chirped zero-area pulse, the excited-state population was measured. The spectral block of $\Delta f(\rm{FWHM})=500$~GHz ($\Delta\omega_2 = 1.9\times 10^{12}$~rad/s) around the resonance was removed in the $4f$-geometry stretcher. In comparison, the corresponding numerical calculation is shown in Fig.~\ref{FigQ}(b). The numerical calculation is based on Eq.~\eqref{S_equation}. Since the experiment was performed under the condition of non-uniform spatial profiles of the laser pulse and the atom cloud, the calculation took into account the spatial averaging effect~\cite{LeeOL2015}, where $\Theta_{\rm Rabi}^{\rm max}$ in the $x$-axis denotes the maximum of the Gaussian distribution of $\Theta_{\rm Rabi}$. The measured populations were calibrated to probabilities by using the first peak of Rabi oscillations as a reference. Within the available measurement region, the experimental result shows that the dark CPR region is surrounded by the bright CPI region, in a good agreement with the theoretical calculation. In addition, we probed the effect of the width of the spectral hole. The frequency chirp was fixed at $c_2= 5\times 10^4$~fs$^2$, and the spectral hole was varied from 200 to 1000~GHz (FWHM). In Fig.~\ref{FigQ}(c), the measured pulse energies (shown with squares) for the first CPI peaks are displayed in comparison with the corresponding numerical calculation (dots). 

We now consider possible applications of the chirped zero-area excitation implicated from the results obtained in this study. 
Since the two-state system interacting with the chirped zero-area pulse can undergo, amid the adiabatic evolution ($z$-rotation) by the main pulse, Rabi-like oscillation due to the non-adiabatic interaction by the hole pulse, an interplay between Rabi-like oscillations and adiabatic evolutions can be made by shaping the spectral width and position of the hole. This approach can be a powerful and alternative control means in selective excitation of, in particular, multi-state systems. In a $V$-type system~\cite{LimPRA2011} of 5S$_{1/2}$, 5P$_{1/2}$, and 5P$_{3/2}$ in rubidium, for example, a numerical simulation (not shown) with chirped zero-area pulses results in over 97\% population of the system driven to either excited states, by simply changing the pulse intensity only. Also, this control method implicates that the laser spatial profile is also useful for position-dependent selective excitations, which can be applied to, for example, atom or ion qubits in spatial arrangements~\cite{BeugnonNP2007, BlattNature2008}.

\section{Conclusion}
We have presented our theoretical and experimental investigation of the two-state system dynamics under the interaction of chirped zero-area pulses. In experiments, we used femto-second laser pulses of which the resonant spectral component was removed and then the pulses were spectrally chirped. These chirped zero-area pulses drove coherent excitations of the resonant 5S$_{1/2}$-5P$_{1/2}$ transition of rubidium atoms that were tightly confined in a MOT within a fraction of the laser spatial profile to ensure uniform laser-atom interactions. The interplay between the adiabatic evolution and Rabi-like oscillations, both of which were simultaneously induced by the chirped zero-area pulses, has been probed. We have shown that the given dynamics can be modeled to the Ramsey-type three-pulse interaction, in a good agreement with the experimental results. The underlying mechanism behind the observed coherent dynamics can be understood based on the nature of the chirped zero-area pulse: The chirped zero-area pulse is a sum of two pulses with different bandwidths, and the broad-band pulse (the main pulse) makes adiabatic evolution even  by a small amount of chirp. However, the narrow-band pulse (the hole pulse) makes either adiabatic or non-adiabatic evolution (thus Rabi oscillation). As a result, the latter switches on or off a Rabi transition amid the adiabatic evolution induced by the former. The result suggests a new design scheme of laser pulse shaping towards selective excitation in multi-state atoms, by embedding a Rabi-like oscillation in adiabatic evolutions.

\begin{acknowledgements}
This research was supported by Samsung Science and Technology Foundation [SSTF-BA1301-12]. The experimental apparatus 
was constructed in part supported by Basic Science Research Program [2013R1A2A2A05005187] through the National Research Foundation of Korea.
\end{acknowledgements}

\end{document}